\documentclass[twocolumn]{aastex63}
\usepackage{amsmath}
\usepackage{color}
\hypersetup{linkcolor=red,citecolor=blue}

\shortauthors{Zhong et al.}

\begin{document}
\title{\textbf{Unveiling the mechanism for the rapid acceleration phase in a solar eruption}}
\correspondingauthor{Ze Zhong}\email{zezhong@sdu.edu.cn}

\author[0000-0001-5483-6047]{Ze Zhong}
\affil{Center for Integrated Research on Space Science, Astronomy, and Physics, Institute of Frontier and Interdisciplinary Science, Shandong University, Qingdao 266237, China\\}
\affil{School of Astronomy and Space Science and Key Laboratory of Modern Astronomy and Astrophysics, Nanjing University, Nanjing 210023, China\\}
\affil{Max Planck Institute for Solar System Research, Justus-von-Liebig-Weg 3, D-37077, G{\"o}ttingen, Germany\\}

\author[0000-0002-9293-8439]{Yang Guo}
\affil{School of Astronomy and Space Science and Key Laboratory of Modern Astronomy and Astrophysics, Nanjing University, Nanjing 210023, China\\}

\author{Thomas Wiegelmann}
\affil{Max Planck Institute for Solar System Research, Justus-von-Liebig-Weg 3, D-37077, G{\"o}ttingen, Germany\\}

\author[0000-0002-4978-4972]{M. D. Ding}
\affil{School of Astronomy and Space Science and Key Laboratory of Modern Astronomy and Astrophysics, Nanjing University, Nanjing 210023, China\\}

\author[0000-0001-6449-8838]{Yao Chen}
\affil{Center for Integrated Research on Space Science, Astronomy, and Physics, Institute of Frontier and Interdisciplinary Science, Shandong University, Qingdao 266237, China\\}

\begin{abstract}
Two major mechanisms have been proposed to drive the solar eruptions: the ideal magnetohydrodynamic instability and the resistive magnetic reconnection. Due to the close coupling and synchronicity of the two mechanisms, it is difficult to identify their respective contribution to solar eruptions, especially to the critical rapid acceleration phase. Here, to shed light on this problem, we conduct a data-driven numerical simulation for the flux rope eruption on 2011 August 4, and quantify the contributions of the upward exhaust of the magnetic reconnection along the flaring current sheet and the work done by the large-scale Lorentz force acting on the flux rope. Major simulation results of the eruption, such as the macroscopic morphology, early kinematics of the flux rope and flare ribbons, match well with the observations. We estimate the energy converted from the magnetic slingshot above the current sheet and the large-scale Lorentz force exerting on the flux rope during the rapid acceleration phase, and find that (1) the work done by the large-scale Lorentz force is about 4.6 times higher than the former, and (2) decreased strapping force generated by the overlying field facilitates the eruption. These results indicate that the large-scale Lorentz force plays a dominant role in the rapid acceleration phase for this eruption.
\end{abstract}


\section{Introduction} \label{sec:intro}
Coronal mass ejections (CMEs) are explosive releases of magnetic energy \citep{2000Forbes}. Their early kinematics usually consist of two distinct phases, a slow rise phase followed by a rapid acceleration phase \citep{2008Schrijver,2020Cheng}. Several mechanisms have been put forward to explain the initiation of the eruption \citep{2018Green}. One category invokes ideal magnetohydrodynamic (MHD) instabilities such as the helical kink instability \citep{1979Hood,2004Torok}, tilt-kink instability \citep{2014Keppens} and double-arc instability \citep{2017Ishiguro}. The other requires a change in the magnetic topology, involving magnetic reconnection. Candidate mechanisms include the tether-cutting reconnection \citep{2001Moore,2021Jiang}, magnetic breakout \citep{1999Antiochos,2012Karpen}, flux emergence \citep{2000Chen}, converging motions \citep{2003Amari}, etc.

For the rapid acceleration phase of solar eruptions, it may exist two underlying mechanisms, i.e., the ideal MHD instability, such as torus instability \citep{2006Kliem,2010Olmedo}, also known as the loss of equilibrium or catastrophe \citep{1978van,1991Forbes} and magnetic reconnection. The loss of equilibrium results from an imbalance between different components of the large-scale Lorentz force while the torus instability occurs if the overlying magnetic field decays fast enough. Both expressions describe the same physical mechanism from two perspectives \citep{2010Demoulin,2014Kliem}. After the force imbalance, the magnetic flux rope (MFR) accelerates to erupt associated with the work done by the large-scale Lorentz force \citep{2006Chen}. The resistive magnetic reconnection occurs in the current sheet between the flare and CME. Upward outflows from the reconnection exhaust region make a primary contribution to accelerate the CME \citep[e.g.,][]{2021Jiang}.

It is difficult to clarify which mechanism plays a leading role in the rapid acceleration phase due to coupling between the above processes \citep{2016Vrsnak}. When magnetic reconnection occurs, it may strengthen the hoop force on the MFR by transferring the overlying flux into the MFR, and reduce the strapping force of the overlying field. Instead, when the force balance is broken, stretched field lines may leave a current sheet behind the MFR, which is favorable for the reconnection to produce outflows. One way to distinguish the two processes is to study whether one of them can drive the solar eruption alone. \citet{2001Shibata} assumed that the MFR is solely affected by the momentum exchange of reconnection outflows in a simplified framework and reproduce typical CME kinematic features. Using three-dimensional (3D) MHD simulations, \citet{2021Jiang} showed that their MFR eruption is mainly driven by the reconnection outflows. 

In contrast to resistive processes, ideal MHD processes have also been confirmed to accelerate a CME alone. \citet{2007Chen} found that the ideal MHD catastrophe can produce the fast CME without magnetic reconnection. \citet{2006Kliem} derived the threshold of the torus instability and reproduced the kinematics of fast and slow CMEs qualitatively. The above researches demonstrate that either of the two processes could drive an eruption alone in the theoretical framework. Nevertheless, their respective contribution remains elusive in real observations. Benefiting from the developed data-driven MHD technology \citep{2016bJiang,2017Leake,2018Hayashi,2019Guo,2021Kaneko,2022Jiang,2022Inoue,2023Chen}, it is possible to disentangle their coupling by quantifying specific contributions, such as reconnection outflows and the work done by the large-scale Lorentz force.

In this letter, we focus on stereoscopic observations involving a single M class flare associated with a CME. By reconstructing the 3D coronal magnetic field, we find an MFR existing before the flare onset. In particular, we perform a data-driven MHD simulation input by the observed time series of the photospheric vector magnetograms and velocities to reproduce the early kinematics of the eruption. Our purpose is to clarify which mechanism dominates the rapid acceleration phase. In Section \ref{s:obs}, we introduce the multi-wavelength observations. The data-driven MHD model is described in Section \ref{s:mhd}. The results are shown in Section \ref{s:results}. The summary and discussion are provided in Section \ref{sec:summary}.

\section{Observations and Data Analysis}\label{s:obs}
\begin{figure*}
\centering
\includegraphics[width=1.0\textwidth]{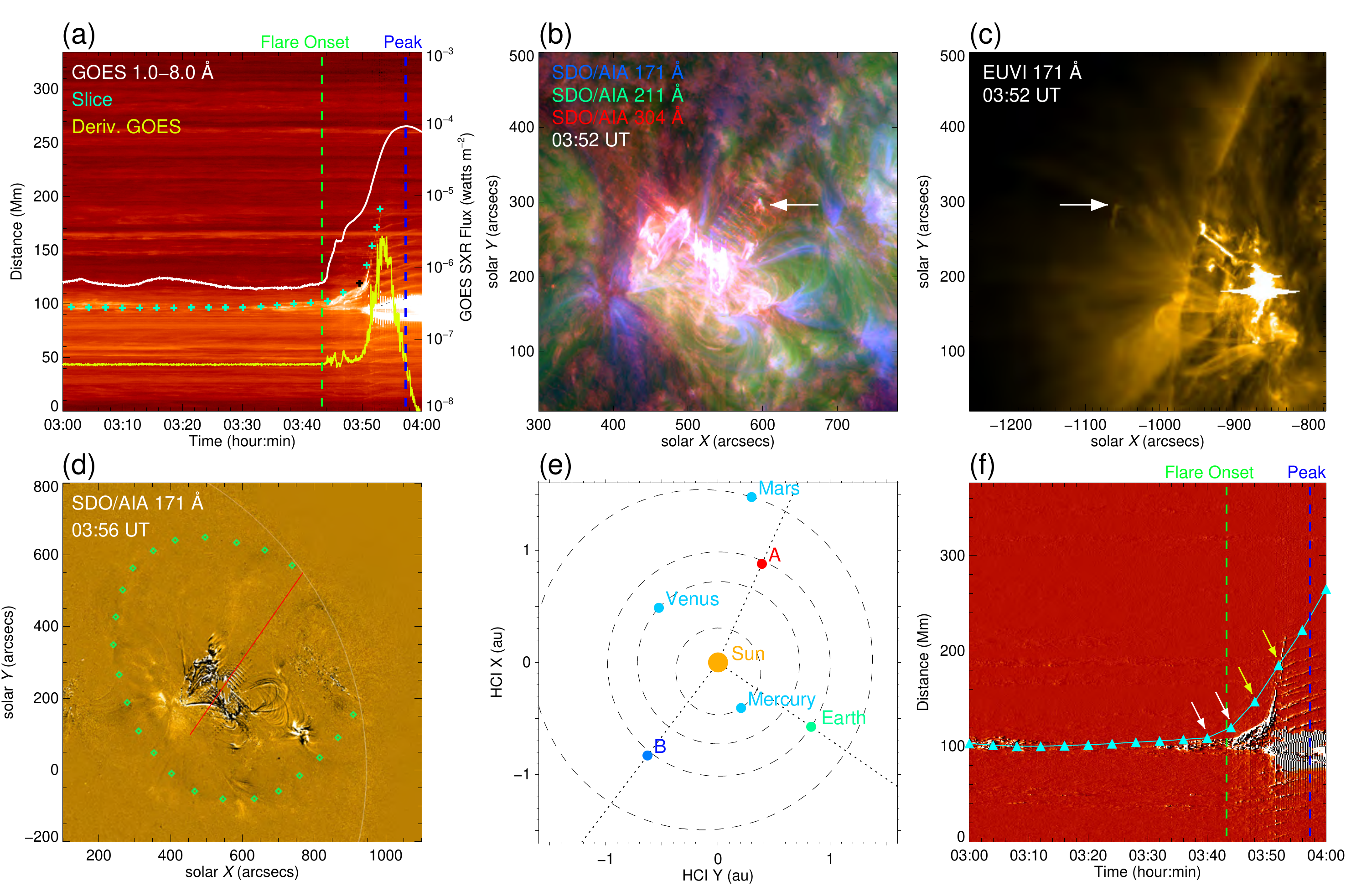}
\caption{
(a) Time-distance diagram of AIA 304 {\AA} images displays the motion of the ejecta. The cyan plus symbols denote the moving front and the black plus marks the turning point. The white and yellow curves show \emph{GOES} soft X-ray flux data of 1.0--8.0 {\AA} and its derivative. The green and blue vertical dashed lines indicate the flare onset and peak times at 03:42 and 03:58 UT, respectively.
(b) Composite images of AIA 171, 211 and 304 {\AA}. The white arrow denotes the ejecta.
(c) The white arrow indicates the same ejecta as that in panel (b), observed by STEREO\_A EUVI 171 {\AA}.
(d) Running difference image of AIA 171 {\AA} image at 03:56 UT. The white arc shows the solar limb. Green diamonds depict the front of the EUV wave. The red oblique line indicates the direction of the eruption.
(e) Positions of STEREO A and B at 03:52 UT in the HCI coordinate system.
(f) Time-distance diagram of AIA 304 {\AA} running difference images after correcting the projection effect. The cyan triangle shows the MFR height measured from the MHD simulation. The white and yellow arrows refer to four moments at 03:40, 03:44, 03:48 and 03:52 UT, respectively. The vertical dashed lines are the same as that in panel (a). An animation of panels (a) and (b) is available. It shows the filament eruption from 03:00 to 04:00 UT in the composite images.
}
\label{fig01}
\end{figure*}

Complex active region (AR) 11261 produced an M9.3-class flare on 2011 August 4. The soft X-ray 1--8 {\AA} flux data of the \textit{Geostationary Operational Environmental Satellite} (\emph{GOES}) show the flare started at 03:42 UT, peaked at 03:58 UT and lasted $\sim$2 hours (Figure \ref{fig01}(a)). According to the observations of the Atmospheric Imaging Assembly \citep[AIA;][]{2012Lemen} on board the Solar Dynamics Observatory \citep[SDO;][]{2012Pesnell} and the Extreme UltraViolet Imager (EUVI) of the Sun Earth Connection Coronal and Heliospheric Investigation \citep[SECCHI;][]{2008Howard} on board the Solar Terrestrial Relations Observatory \citep[STEREO;][]{2008Kaiser}, an arc-shaped filament situated along the polarity inversion line (PIL) before the flare onset. Subsequently, the flare occurred, and the brightening plasma ejected quickly from the AR (white arrows in Figures \ref{fig01}(b) and \ref{fig01}(c)) with much emission in the composite images. After that, a series of post-flare loops formed as seen in 171 {\AA} and a large EUV wave passed through the solar disk (green diamonds in Figure \ref{fig01}(d)). According to the coronagraph data observed by C2 and C3 of the Large Angle and Spectrometric Coronagraph \citep[LASCO;][]{1995Brueckner}, A large CME was then detected in the field of view (FOV) of C2 at $\sim$04:12 UT, and in the FOV of C3 at $\sim$04:30 UT. The early kinematics of the solar eruption was well observed by these tandem of instruments.

To quantify the kinematics of the ejecta, we make a time-distance plot in Figure \ref{fig01}(a) along the artificial slice (red oblique line in Figure \ref{fig01}(d)) and measure the moving front as denoted by the cyan pluses. Before the flare onset, the filament kept relatively stationary (Animation of Figure \ref{fig01}(b)). At 03:42 UT, the flare occurred, then the ejecta accelerated slowly with a linear speed, 46.6 km s$^{-1}$. After the turning point (black plus), the speed of the moving front suddenly increased to 330.1 km s$^{-1}$. Five minutes later, the ejecta left the FOV. We also calculate the derivative of \emph{GOES} soft X-ray flux data as shown by the yellow curve, which is usually assumed to satisfy the Neupert effect \citep{1968Neupert}. Its tremendous change around the turning point matches well with the kinematics of the ejecta. 

Combined with the EUV data observed by the EUVI and AIA, we reconstruct the 3D trajectory of the ejecta. Figure \ref{fig01}(e) shows the positions of STEREO A and B at 03:52 UT in the Heliocentric Inertial (HCI) coordinate system. The separation angle between STEREO A and SDO is about $99^\circ$. The same ejecta in the AIA perspective corresponds to the northeast direction in the FOV of EUVI as shown in Figures \ref{fig01}(b) and \ref{fig01}(c). By linear fitting different ejecta positions in a series of time, we obtain the corrected time-height trajectory of the ejecta as shown in Figure \ref{fig01}(f).

\section{Numerical Modeling Method}\label{s:mhd}
We use an MHD model under the zero-$\beta$ assumption to simulate the evolution of the MFR. The model is largely the same as our previous study \citep{2021Zhong}. The MHD equations are numerically solved in the Cartesian coordinate system by the open source message passing interface adaptive mesh refinement versatile advection code \citep[MPI-AMRVAC;][]{2012Keppens,2014Porth,2018Xia,2021Keppens}. We choose a three-step Runge$-$Kutta method for the time integration. For the spatial discretization, we adopt a finite-volume scheme setup combining the Harten$-$Lax$-$van Leer solver \citep{1983Harten} with a third-order Koren slope limiter \citep{1993Koren}. We use a 3D uniform mesh, with $224 \times 192 \times 192$ cells in the $x$, $y$ and $z$ directions, corresponding to the full computation domain with $330 \times 283 \times 283$ $\mathrm{Mm^3}$. 

\begin{figure*}
\centering
\includegraphics[width=1.0\textwidth]{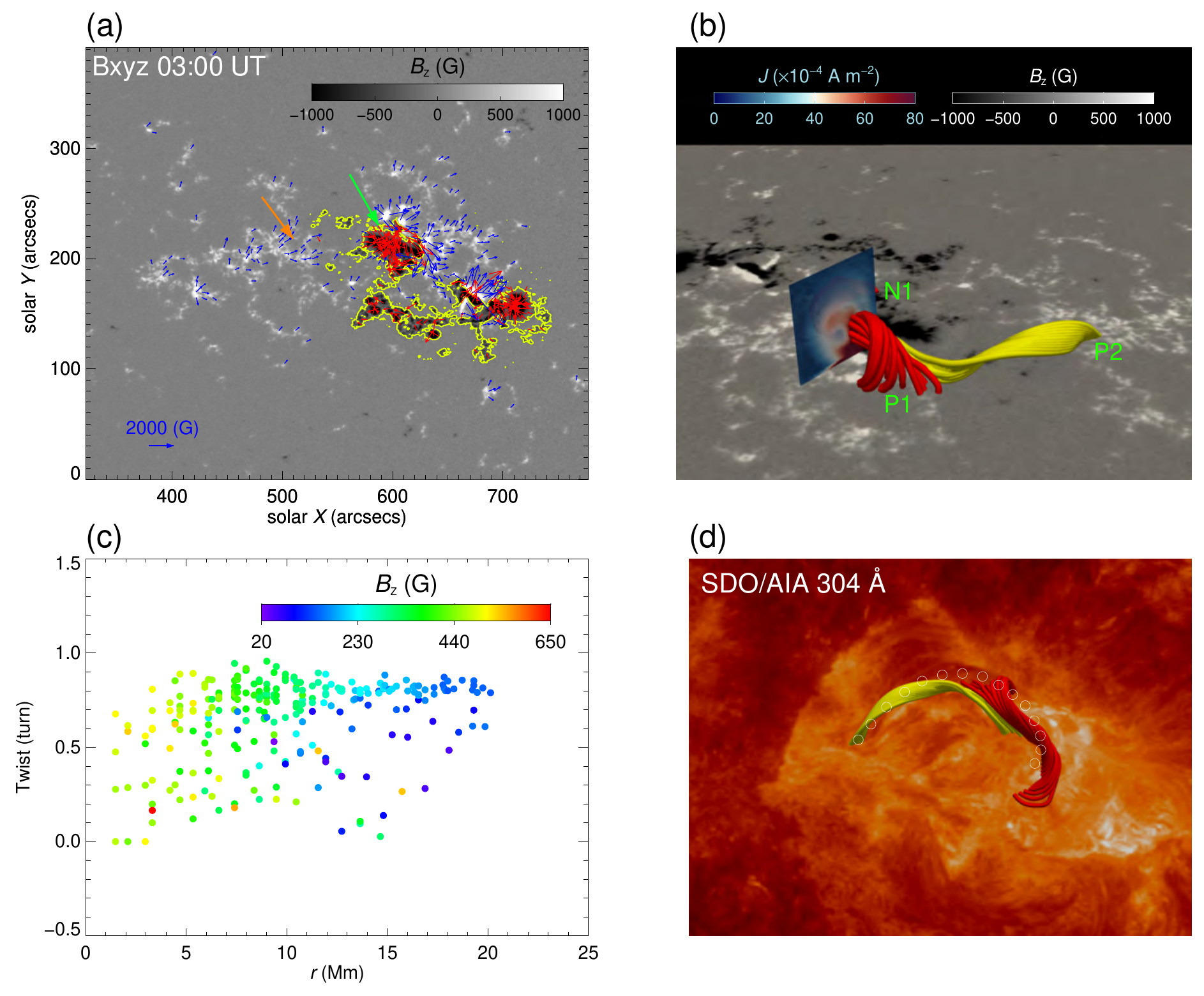}
\caption{
(a) Photospheric vector magnetograms at 03:00 UT with the FOV of $456\arcsec \times 392\arcsec$. The red and blue arrows represent the transverse fields in negative and positive polarities, respectively. The yellow contour shows the $B_z$ with a level of $-$100 G. 
(b) NLFFF extrapolated from the photospheric vector magnetograms at 03:00 UT. The red and yellow lines depict the MFR with three polarities N1, P1 and P2, at the location indicated by green and orange arrows in panel (a), respectively. The semi-transparent vertical slice across the MFR axis displays the distribution of the total electric current density.
(c) The distribution of the twist number as a function of the distance to the MFR axis, $r$. Different colors indicate the strength of $B_z$ at footpoints, which are measured in the positive polarity.
(d) The MFR from a top view, overlaid on the AIA 304 {\AA} image. The filament is outlined by white circles.
}
\label{fig02}
\end{figure*}

We use the magneto-frictional method \citep{2016Guo} to reconstruct the nonlinear force-free field (NLFFF) for the initial magnetic configuration based on the photospheric vector magnetograms at 03:00 UT (Figure \ref{fig02}(a)), taken by the Helioseismic and Magnetic Imager \citep[HMI;][]{2012Scherrer} on board \emph{SDO}. We correct the projection effect \citep{1990Gary} and remove the net Lorentz force and torque by the preprocessing \citep{2006Wiegelmann} to ensure the force-freeness and torque-freeness. We combine the preprocessed vector magnetograms and the potential field as an initial non-force-free field to relax to a force-free state. The final NLFFF is close to a force-free and divergence-free state, which can be evaluated by two metrics \citep{2000Wheatland}. One is the force-freeness metric, defined by the current weighted average of the sine of the angle between the current density $\textbf{\emph{J}}$ and magnetic field $\textbf{\emph{B}}$. The other is the divergence-freeness metric, determined by the weighted average of the fractional magnetic flux increase. The force-freeness metric in our study is $\sim 0.16$, equivalent to $9.2^\circ$ for the angle between $\textbf{\emph{J}}$ and $\textbf{\emph{B}}$, and the latter is $\sim 2.2\times 10^{-4}$. They are similar to the previous NLFFF using the magneto-frictional method \citep[e.g.,][]{2022He} and optimization method \citep[e.g.,][]{2021Jing}. 

Figure \ref{fig02}(b) shows the initial configuration including an MFR in the low atmosphere, with a height of 11 Mm. The MFR center has an intense current density (8 mA m$^{-2}$), about three times higher than the edge, as shown in the vertical slice. The main body of the MFR connects the negative polarity N1 and the positive polarity P1, while partial footpoints extend to the eastern positive polarity P2. We calculate the twist number of the MFR by a geometric method \citep{2017Guo} as shown in Figure \ref{fig02}(c). The max twist number is approximately 1 turn. We also compare the morphology of the MFR and pre-eruption filament in Figure \ref{fig02}(d). The path and footpoints of the filament are overall consistent with the MFR.

The initial atmosphere is a plane-parallel model, where the auxiliary temperature is the same as previous studies \citep{2019Guo,2021Zhong,2021Guo}. The density is solved by the hydrostatic equilibrium, $\emph{d} (\rho T) / \emph{d} h = -\rho g$. In our setup, $\rho$, is $2.3 \times 10^{-7}$ g cm$^{-3}$ on the bottom boundary, and it decreases to $3.1 \times 10^{-15}$ g cm$^{-3}$ on the top boundary. Note that the temperature is only used to solve the density distribution, and it does not appear in the zero-$\beta$ equations.

In boundary conditions, we set all six boundaries as their initial values and keep them unchanged for the density. The magnetic field in the bottom consists of a time-series of vector magnetograms, from 03:00 to 04:24 UT with an interval of 12 minutes. All magnetograms are aligned to 03:00 UT and preprocessed as same as that in the extrapolation. We assign these magnetograms to the inner ghost layer on the bottom boundary and make a second-order constant value extrapolation to the outer ghost layer. The derived velocities are calculated by the Differential Affine Velocity Estimator for Vector Magnetograms \citep[DAVE4VM;][]{2008Schuck}. We set them to the inner ghost layer on the bottom boundary and make a constant value extrapolation to the outer ghost layer. For top and side boundaries, the magnetic fields are given by the constant value extrapolation, while the velocities are set to be zero.

\section{Results}\label{s:results}
\subsection{Simulation Results Compared with Observations}\label{ss:evolution}
\begin{figure*}
\centering
\includegraphics[width=1.0\textwidth]{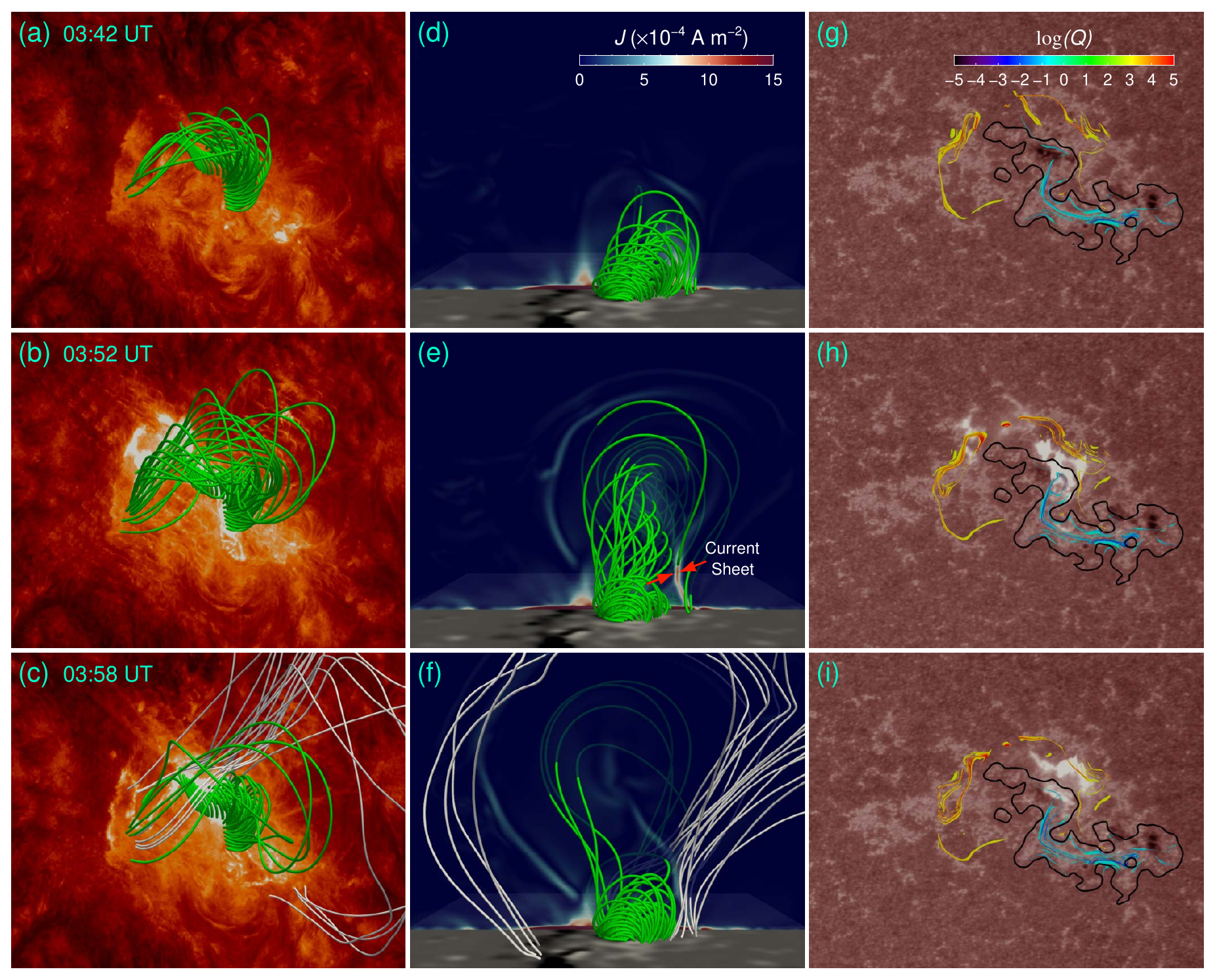}
\caption{Three typical snapshots display the morphology of the flare at 03:42, 03:52 and 03:58 UT corresponding to the flare onset, impulse and peak times.
(a)--(c) Magnetic field lines overlaid on the AIA 304 image. The green and gray lines represent the MFR and overlying fields, respectively.
(d)--(f) A side view along the $x$-axis of the magnetic field. The vertical slice displays the distribution of the total current density. The bottom boundary is photospheric magnetogram. There is an extended current sheet in panel (e) and two red arrows indicate the local reconnection inflows.
(g)--(i) Comparison between the QSLs on the bottom and the AIA 1700 {\AA} emission, overlaid by the contour of $B_z$ with a level of $-$20 G. The QSLs with $|\mathrm{log}(Q)| > 3$, are signed with positive and negative polarities. An animation of this ﬁgure is available. It displays the evolution of the magnetic field compared to AIA 304 and 1700 {\AA} images from 03:00 to 04:20 UT.
}
\label{fig03}
\end{figure*}

From two different perspectives, Figures \ref{fig03}(a) and \ref{fig03}(d) show the morphology of the MFR at 03:42 UT corresponding to the flare onset. Distinct from 03:00 UT, a part of footpoints in positive polarity moves from polarities P1 to P2, although the height of the MFR is almost unchanged. Then, the MFR becomes unstable and starts to rise. In the flare impulsive phase (Figures \ref{fig03}(b) and \ref{fig03}(e)), the MFR undergoes a rapid eruption, and a current sheet is stretched below it. At 03:58 UT (Figures \ref{fig03}(c) and \ref{fig03}(f)), close to the flare peak, the MFR reaches to about 175 Mm, and the volume of it becomes very large due to the continuous expansion.

To compare the magnetic field with the observations directly, we back-project the magnetic field to the heliocentic coordinate system and superimpose them on AIA 304 {\AA} images (Figures \ref{fig03}(a)$-\ref{fig03}$(c)). We delineate typical lines with different colors to distinguish the MFR (green) and overlying fields (gray). Two aspects of the MHD simulation are consistent with the observations. In terms of the time evolution, the flare onset coincides with the time when the MFR begins to rise, and the flare peak associates with the rapid rise of the MFR. For morphology, the MFR trajectory is basically consistent with the eruptive filament, both along the northwest. The footprints of field lines also match well with the flare ribbons in 304 {\AA} images. 

We calculate the bottom quasi-separatrix layers (QSLs) to trace the change of the magnetic topology qualitatively. The QSLs are generally considered as positions with strong gradients of the magnetic connectivity \citep{2002Titov}. At these regions, the magnetic reconnection may occur and produce the brightening of the flare ribbons. We identify the QSLs by calculating the squashing factor $Q$ on the photosphere and superimpose the signed QSLs with $\mathrm{log}(Q) > 3$ on the 1700 {\AA} images (Figures \ref{fig03}(g)$-$\ref{fig03}(i)) at three typical moments corresponding to flare onset, flare impulsive phase and flare peak. We find that the evolution of the QSLs is overall consistent well with the flare ribbons. Especially after the flare onset, the QSLs begin to separate on either side of the PIL. It illustrates that our simulation achieves relatively high fidelity in the evolution of the magnetic topology.

We also derive the time-height profile of the MFR (cyan triangles) from the simulation to compare with the corrected time-distance profile as shown in Figure \ref{fig01}(f). Both manifest a slow rise phase followed by a rapid acceleration phase. Before the flare onset, the MFR undergoes a stable evolution in 40 minutes. After that, the MFR has a slow rise with a uniform speed, 45.8 km s$^{-1}$, almost consistent with the observed speed. Then, after the turning point, the speed of the MFR increases sharply, approaches to 179.2 km s$^{-1}$, about 41.1\% of the observed speed. By comparing the two stages, we find that the simulation basically reproduces the observed features, especially for the transition from slow to fast acceleration phases.

\subsection{The Role of Reconnection Flow}\label{ss:RF}
\begin{figure*}
\centering
\includegraphics[width=1.0\textwidth]{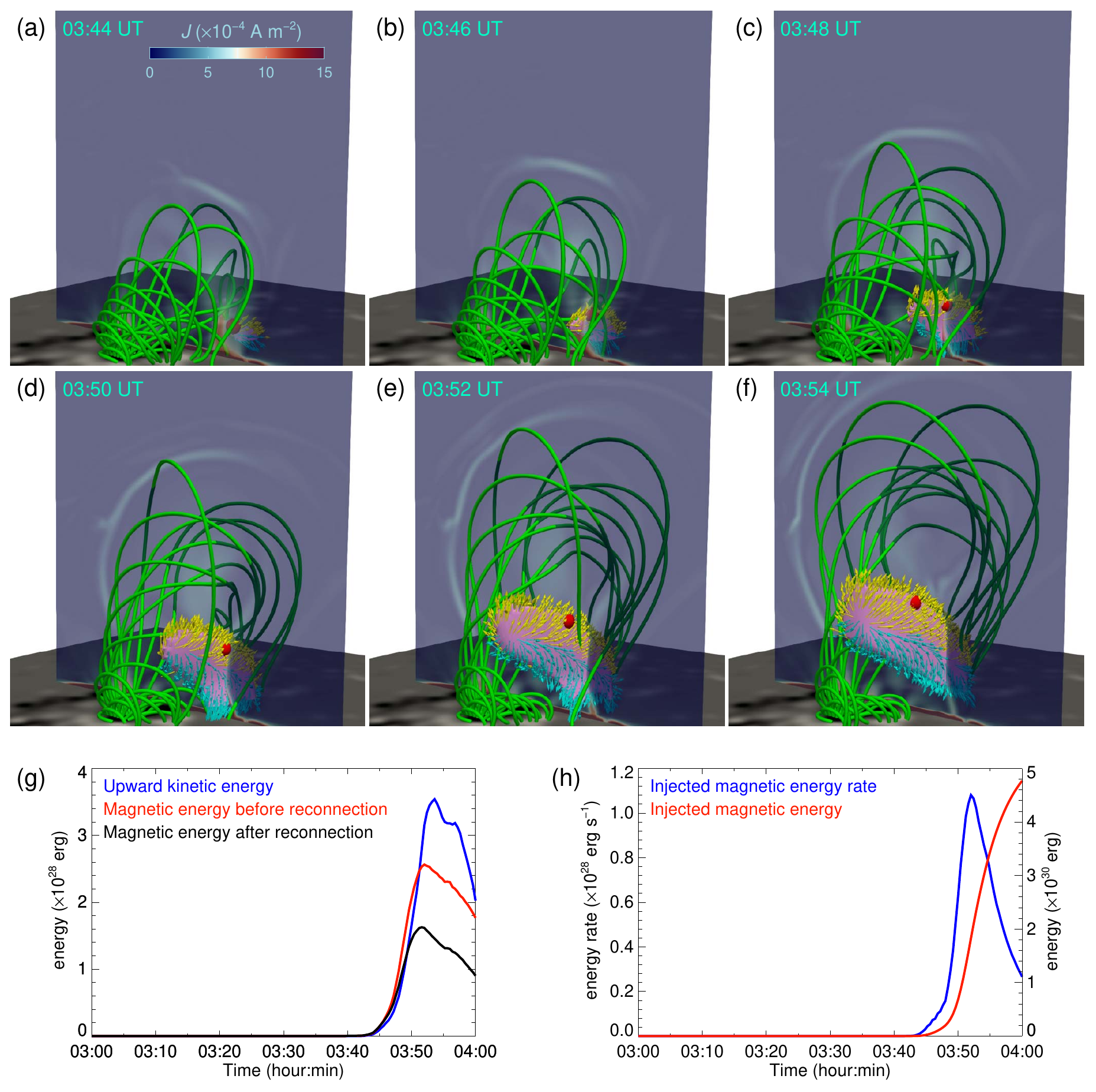}
\caption{
(a)--(f) Six snapshots depict the development of the 3D current sheet (pink isosurface) at 03:44, 03:46, 03:48, 03:50, 03:52, and 03:54 UT. The yellow and cyan arrows represent the reconnection outflows with two opposite directions, respectively. The red dot shows the fastest point of upward outflows. The green field lines represent the main body of the MFR. The vertical semi-transparent slice displays the distribution of the total current density while the bottom boundary denotes the photospheric magnetogram.
(g) The blue, red and black lines show the temporal evolution of the upward kinematic energy, magnetic energy before and after the reconnection in the defined 3D current sheet, respectively.
(h) Total magnetic energy (red) injected into the 3D current sheet with time and its derivative (blue).
}
\label{fig04}
\end{figure*}

Reconnection outflows are recognized as an important mechanism to accelerate the MFR \citep[e.g.,][]{2021Jiang}. We focus on the outflow site and identify it by defining the location of the current sheet. Analogous to \citet{2006Gibson} and \citet{2016Jiang}, here the current sheet is defined as the region with the ratio of $J/B=|\nabla \times \textbf{\emph{B}}|/|\mu_0 \textbf{\emph{B}}|$ greater than $0.35/\mu_0 \Delta X$, where $J$ is current magnitude, $B$ is magnetic strength, and $\Delta X$ is the grid resolution. Such a definition is convenient to identify regions of high resistivity where magnetic reconnection occurs. Note that the current sheet under this definition has a finite volume with a thickness of $\sim$3$\Delta X$, rather than an infinitely thin structure in the standard flare model. Figures \ref{fig04}(a)$-$\ref{fig04}(f) show a full 3D evolution of magnetic reconnection after the flare onset. With the MFR ascending, the 3D current sheet (pink isosurface) expands and the upward outflows (yellow arrows) push the MFR. The fastest point (red) of upward outflows also moves in space.

We calculate three quantities in the defined 3D current sheet including upward kinetic energy, magnetic energy before and after reconnection, to evaluate the role of upward outflows, as shown in Figure \ref{fig04}(g). Here, we use a fixed moment 03:40 UT to approximately calculate the magnetic energy before magnetic reconnection in the dynamic 3D current sheet (red curve). The magnetic energy after magnetic reconnection is contributed by the local magnetic field. Note that the upward kinetic energy depends on the velocity of the fluid element and the space it occupies at different time, while the magnetic energy before and after reconnection is associated with the volumetric variation of the current sheet. The evolution of the three quantities displays a similar tendency, i.e., divides into three stages. Before the flare onset, all of quantities are zero due to no current sheet. With the MFR rising, three quantities increase impulsively, reach the peak value close to 03:54 UT, and decrease gradually in the end. The maxima of the three quantities are in the same order of magnitude. 

The magnetic energy injecting to the 3D current sheet is defined as:
\begin{equation}
E=-\int_0^t \oint_S \textbf{\emph{P}} \cdot \emph{d} \textbf{\emph{S}} \mathrm{d} t = - \int_0^t \int_V \nabla \cdot \textbf{\emph{P}}~\mathrm{d} V \mathrm{d} t 
\end{equation}
where $\textbf{\emph{P}}=\frac{1}{\mu_0}\textbf{\emph{B}}\times(\textbf{\emph{v}}\times\textbf{\emph{B}})$ is Poynting flux injected from the boundary, $V$ is the region of the defined current sheet bounded by a closed surface $S$, with $J/B>0.35/\mu_0 \Delta X$. Figure \ref{fig04}(h) shows the total magnetic energy (red) and its time derivative (blue) injected into the 3D current sheet, respectively. The blue curve has the similar trend compared to the first three quantities. Note that the total injected magnetic energy is nearly hundred times higher than the magnetic energy after reconnection. It suggests that most of the magnetic energy injected into the current sheet is released.

\subsection{Lorentz Force Analysis}\label{ss:LF}
\begin{figure*}
\centering
\includegraphics[width=0.72\textwidth]{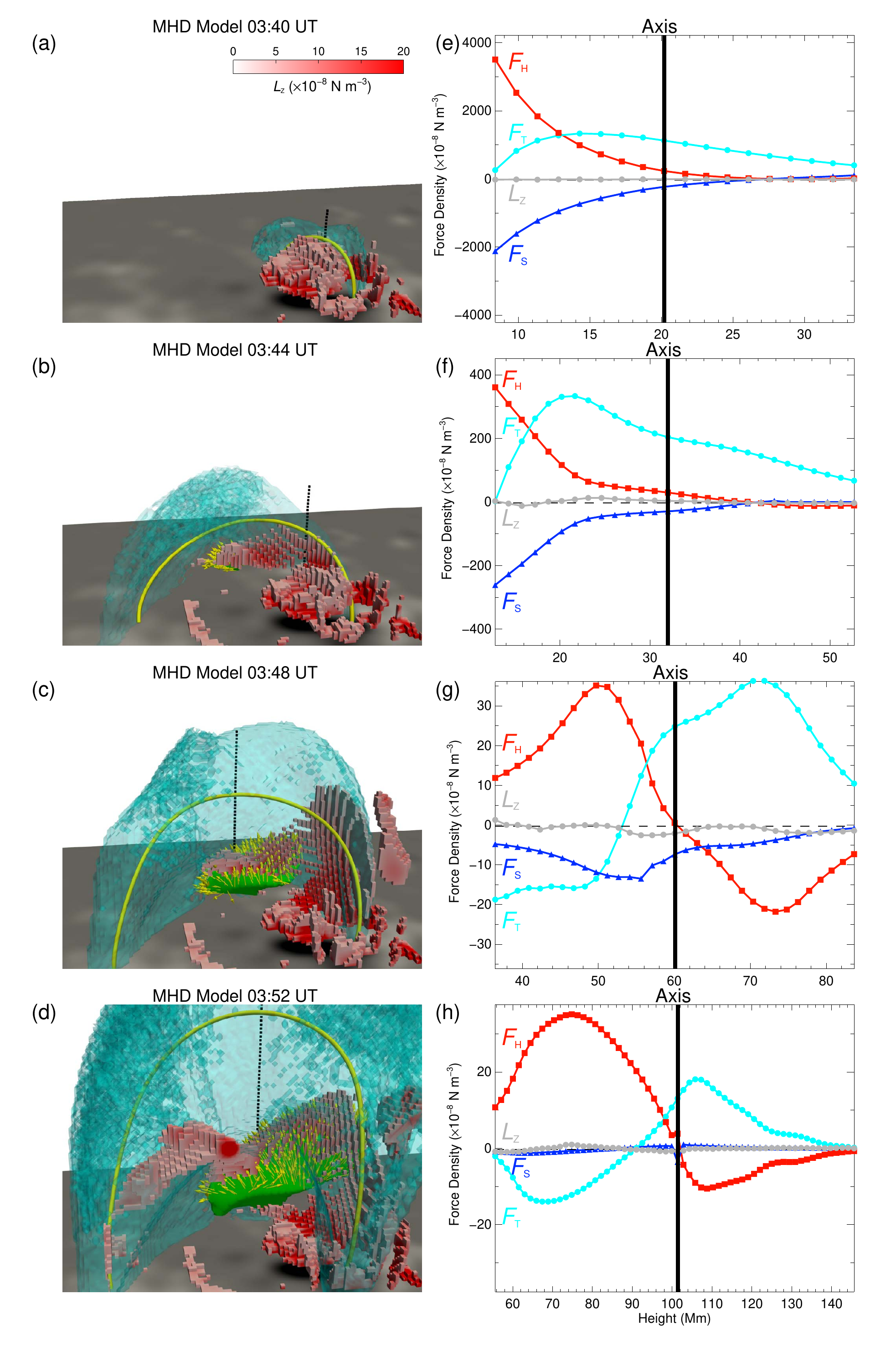}
\caption{Four typical moments display the Lorentz force at 03:40, 03:44, 03:48 and 03:52 UT. The two moments are before the turning point and the other two in behind.
(a)--(d) 3D distribution of the upward $L_z$. The cyan semi-transparent isosurface outlines the MFR. The yellow line denotes the MFR axis. The green isosurface shows the 3D current sheet, with upward (yellow) outflows. The black dotted line covers a height range from 8.4 to 33.5 Mm in panel (a), 12.8 to 52.7 Mm in panel (b), 36.4 to 83.7 Mm in panel (c) and 55.6 to 145.6 Mm in panel (d). The bottom boundary displays the photospheric magnetogram.
(e)--(h) Distribution of the vertical Lorentz force ($L_z$, gray), hoop force ($F_{\mathrm{H}}$, red), strapping force ($F_{\mathrm{S}}$, blue) and tension force ($F_{\mathrm{T}}$, cyan), along the black dotted lines in panels (a)--(d), respectively. The vertical black line shows the location of the MFR axis. All abscissas are different due to the rising axis.
} 
\label{fig05}
\end{figure*}

The force imbalance originates from the competition between different components of the large-scale Lorentz force. We first check the overview of the large-scale Lorentz force by calculating the vertical component, $L_z=\textbf{\emph{e}}_z \cdot (\textbf{\emph{J}} \times\textbf{\emph{B}})$. Figures \ref{fig05}(a)$-$\ref{fig05}(d) display the 3D distribution of upward $L_z$ in four typical moments at 03:40, 03:44, 03:48 and 03:52 UT (four arrows in Figure \ref{fig01}(f)), in which two of them are before and the other two are after the turning point. With the MFR ascending, the 3D current sheet expands and the region with upward outflows (yellow arrows) contributes a part of upward $L_z$. Then we decompose $L_z$ to explore which component makes a major contribution. Followed by \citet{2021Zhong}, we investigate three force components by combinations of the electric current and magnetic field components, including the hoop, tension and strapping forces as shown in Figures \ref{fig05}(e)$-$\ref{fig05}(h). The hoop force (red) always promotes the MFR rising, while the strapping force (blue) constrains the MFR. The tension force (cyan) is variable, initially acting as a driving force, to be a confining force in the end. The relative magnitude of force components is also highly variable. Especially for the strapping force, it becomes smaller after the turning point, resulting in weak constraint further to facilitate the eruption.

We also calculate the work done by the Lorentz force to estimate the magnitude of magnetic energy release both in the region of the 3D current sheet and MFR. The region of the MFR is limited by the QSL boundary. The work is defined as: 
\begin{equation}
\begin{aligned}
W&=\int_0^t \int_V \textbf{\emph{J}} \cdot \textbf{\emph{E}}~\mathrm{d} V \mathrm{d} t=\int_0^t \int_V \textbf{\emph{J}} \cdot (-\textbf{\emph{v}} \times \textbf{\emph{B}})~\mathrm{d} V \mathrm{d} t \\
&=\int_0^t  \int_{V,V_z > 0,L_z > 0} (\textbf{\emph{J}} \times \textbf{\emph{B}})_z \cdot \textbf{\emph{v}}_z~\mathrm{d} V \mathrm{d} t
\end{aligned}
\end{equation}
where $\textbf{\emph{J}}$ is electric current density and $\textbf{\emph{E}}$ is electric field. Note that the work is integrated in the region with upward motion and upward Lorentz force. The work done by the vertical component of the Lorentz force during the eruption is roughly estimated to be about $9.58 \times 10^{29}$ erg and $4.40 \times 10^{30}$ erg in the region of the 3D current sheet and MFR, respectively. The former represents the slingshot effect while the latter reflects the impact of the large-scale Lorentz force acting on the MFR. It indicates that the work done by the large-scale Lorentz force serves as a major mechanism to accelerate the MFR. In addition, we note that the magnetic energy injected into the current sheet is $4.78 \times 10^{30}$ erg, which mostly released by reconnection. It converts into the kinetic energy of upflows and downflows, free energy in the magnetic field, and other energy forms, which illustrates the important role of magnetic reconnection.

\section{Summary and Discussion} \label{sec:summary}
We investigated a solar eruption on 2011 August 4 using the data-driven MHD model with a time series of SDO/HMI magnetograms as the boundary condition input. The initial condition is an approximately force-free magnetic field, extrapolated from the photospheric magnetograms at 03:00 UT, about 42 minutes before the flare onset. We found that the modeled results basically reproduce four aspects including the macroscopic morphology, 3D eruptive trajectory, MFR early kinematics and simulated QSLs, compared to the SDO/AIA multi-wavelength observations.

We analyzed the role of reconnection outflows and the large-scale Lorentz force for the rapid acceleration. We first assessed the impact of the reconnection outflows. In the standard scenario, when magnetic reconnection occurs, the current sheet is stretched continuously, and lots of reconnection outflows will promote the MFR to accelerate \citep[e.g.,][]{2021Jiang}. In our simulation, the growth of the current sheet stops near the flare peak. Although many plasmas flow into the current sheet after the flare onset, a small amount of magnetic energy has been converted to the kinetic energy. Second, we calculated the work done by the large-scale Lorentz force exerting on the MFR and magnetic slingshot above the current sheet. The former is 4.6 times times higher than the latter, which illustrates that at least for the event simulated here the large-scale Lorentz force plays a major role in the rapid acceleration phase. By further decomposing the Lorentz force, we confirm that the reduced strapping force of the overlying field causes the imbalance, and the resultant upward Lorentz force propels the MFR to accelerate.

Although we have evaluated the effect of reconnection outflows quantitatively, it is difficult to quantify the topology change induced by reconnection. In the real situation, the MFR and overlying fields are highly dynamic. We need to further distinguish the newly added flux from shear arcade, the reduced overlying fields and the flux transformation in the MFR due to the internal reconnection. More importantly, the topology change affects the distribution of the large scale Lorentz force on the MFR directly. Its subsequent effect is reflected in the change of Lorentz force. Therefore, we do not preclude the role of magnetic reconnection. On the contrary, the solar eruption can never erupt without magnetic reconnection.

We note that the speed of the MFR in the second stage is slower and the turning point is 6 minutes earlier than that in the observation. One can see an obvious filament in the AIA observations before the flare onset. \citet{2021Guo} adopted two density values in the two stages to mimic the drainage of the filament because the zero-$\beta$ approximation cannot reveal the thermodynamics. It is conceivable that the gravity may delay the MFR lifting and the speed of the MFR may increase a lot when the mass density decreases. Currently, no additional forces can compensate the non-vanishing Lorentz force in the low atmosphere. The energy conversion is also lacking under the zero-$\beta$ assumption. The total magnetic energy injected into the current sheet is primarily released by the work done. Other forms of energy transfer such as conduction and radiation are currently neglected, which requires us to perform a full data-driven MHD model including the thermodynamics \citep{2022Jiang}.

Our simulation provides a possible interpretation for the rapid acceleration in a real solar eruption. We do not preclude other mechanisms due to the intrinsic complexity, even vary from event to event. For example, \citet{2021Jiang} revived the role of the reconnection outflows in the 3D tether cutting model. They considered that the MFR is mainly driven by the upward magnetic tension of the newly reconnected field lines. Magnetic reconnection could strength the hoop force of the MFR \citep{2016Vrsnak}. The increased hoop force also provides higher acceleration and prolongs the late phase of the CME. Additionally, our study does not involve the triggering mechanism, which also has highly complexity (e.g., see review by \citealt{2018Green}). Finally, the driving mechanism of solar eruptions is still a complicated issue due to coupling of various physical factors. We expect to explore it by including more physics and inputting observational data with a higher spatio-temporal resolution to future data-driven MHD models.

\acknowledgments
The authors are grateful to the anonymous referee for the constructive comments and thank the SDO/AIA, HMI and STEREO/SECCHI consortia for supplying the data.
SDO is a mission of NASA's Living With a Star Program. STEREO is the third mission in NASA's Solar Terrestrial Probes program.
Z.Z., Y.G., M.D.D., and Y.C. were supported by the National Key Research and Development Program of China (2022YFF0503004), NSFC (11973031, 11790303, and 11733003) and China Postdoctoral Science Foundation (2022M711931). T.W. was supported by DLR-grant 50 OC 2101.
We acknowledge the computational resources provided by the HPC-cluster in Max Planck Institute for Solar System Research, the cluster system of the High Performance Computing Center in Nanjing University and the Beijing Super Cloud Computing Center (BSCC, URL: http://www.blsc.cn).

\bibliographystyle{aasjournal}
\bibliography{reference}

\end{document}